\documentclass[%
 reprint, 
 amsmath,amssymb,
 aps, 
prb,
showkeys
]{revtex4-2}
\setcitestyle{numbers,sort&compress=false}
\usepackage{multirow}
\usepackage{xcolor}
\usepackage{lineno,hyperref}
\usepackage{gensymb}
\usepackage{tabularx}
\hypersetup{
    colorlinks=true,
    linkcolor=blue,
    bookmarksnumbered,
    citecolor=blue,
    urlcolor=blue,
    bookmarksopen=true
}
\usepackage{graphicx}
\usepackage{dcolumn}
\usepackage{bm}
\usepackage{siunitx}
\usepackage{comment}

\begin{document}

\preprint{APS/123-QED}

\title{Electronic States, Spin-Orbit Coupling and Magnetism in Germanium 60° Dislocations}

\author{Veronica Regazzoni}
\author{Fabrizio Rovaris}
\author{Anna Marzegalli}
\author{Francesco Montalenti}
\author{Emilio Scalise}
\email{emilio.scalise@unimib.it}

\affiliation{Department of Materials Science, University of Milano-Bicocca, Via R. Cozzi 55, I-20125 Milano, Italy}

\begin{abstract}
Defects in semiconductors have recently attracted renewed interest owing to their potential in novel quantum applications. Here we investigate the electronic and magnetic properties induced by 60° dislocations in Ge. Using large-scale DFT calculations, we determine the band structure for both the shuffle and glide sets in their lowest-energy configurations.  
The band structure for the shuffle set reveals defect-induced dispersive bands localized within the band gap near the $\Gamma$ point, whereas for the glide set, we observe strong overlap with the conduction band. Defect-induced band splitting evident away from $\Gamma$ reveals Rashba-Dresselhaus spin-orbit coupling, an effect previously reported only for screw dislocations. Remarkably, we find evidence that specific dislocation arrangements can stabilize antiferromagnetic ordering with sizable local magnetic moments and considerable exchange splitting between opposite spin states. These results uncover rich physics in Ge dislocations through the combination of spin-orbit coupling and magnetic ordering, potentially enabling novel defect-based functionalities in Ge devices.

\end{abstract}


\maketitle

\section{INTRODUCTION}

The rise of quantum technologies in recent years has strengthened the key role of germanium (Ge) in the field of semiconductor research~\cite{scappucci2021germanium,KochNPJQuantum2025,SHIMURA2024108231}.
Ge is typically grown heteroepitaxially on lattice-mismatched silicon substrates, resulting in the accumulation of elastic energy and injection of line defects, known as dislocations, during the strain relaxation process~\cite{matthews1974defects, people1985calculation}. For a long time, these defects were merely considered detrimental but, recently, defect engineering has been explored for possible applications~\cite{ivady2019stabilization, barragan2024assessing, steele2025superconductivity}. The atomic structure and chemical composition of dislocation cores are critical determinants of material functionality, as they can introduce localized states within the bandgap that directly influence carrier transport, recombination dynamics, and optical absorption spectra. Dislocations show striking similarities to low-dimensional quantum systems: their structural stability, combined with tunable electronic and spin properties, positions them as potential building blocks for emerging quantum and spintronic architectures~\cite{polat2023dislocations, krivobok2018two, li2023dislocation, liang2023field, Zhang2026}.

The dislocations produced during the strain relaxation of epitaxial Ge layers operate on the $\{111\}\langle 110\rangle$ slip system~\cite{fitzgerald1989properties, hirth1983theory, humble1978plastic}. In particular, 60° dislocations play a prominent role, as they typically dominate the strain relaxation process. These defects have been observed and extensively characterized by experimental investigations both in constant-composition~\cite{MooneyMSER1996, Bolkhovityanov_2012,PhysRevApplied.10.054067,marzegalliprb2013}, in compositionally-graded~\cite{LeGouesPRL1991, WANG20011599} and in more complex heterostructured SiGe layers~\cite{FitzgerladPssa1999, SkibitzkiPRM2020, ArroyoSciptaMat2019, RovarisPRM2017}. 
Such dislocations belong to either glide or shuffle sets, which differ in bonding configurations and interplanar spacing, with the glide (shuffle) set consisting of closely-spaced (widely-spaced) $\{111\}$ planes. Their distinct core reconstructions have a strong influence on the stability and electronic properties of the dislocations, motivating a focused theoretical investigation of 60° dislocations in both glide and shuffle configurations.

The quasi one-dimensional (1D) nature of dislocations and their long-range elastic interactions require large-scale simulation cells and realistic Hamiltonians to accurately model the dislocation-induced electronic states. In principle, \textit{ab-initio} techniques such as density functional theory (DFT) are well suited for this purpose. However, standard approximations such as the local-density approximation (LDA) and the generalized gradient approximation (GGA) often fail to accurately describe the electronic bandgap of semiconductors. In fact, bulk Ge is predicted to be metallic at the LDA and GGA levels, necessitating the use of more advanced exchange-correlation functionals such as hybrid functionals (e.g., HSE06~\cite{heyd2003hybrid, peralta2006spin}) or meta-GGA functionals (e.g., mBJ~\cite{becke2006simple, tran2007band, tran2009accurate}). Although these methods are computationally more demanding, they provide bandgap values closer to the experimental measurements and are essential for modeling the electronic properties of dislocations~\cite{Scalise2020, Fadaly2021}.

In this study, we investigate the structural stability and electronic properties of the 60° perfect dislocations in Ge, considering both glide and shuffle sets. Using high-level exchange-correlation functionals and advanced atomistic modeling techniques, we aim to understand the impact of these dislocations on the electronic properties of Ge. Additionally, we investigate spin-orbit coupling effects in 60° perfect dislocations, which have been recently reported for screw dislocations in Ge~\cite{hu2018ubiquitous}, and we discuss the possibility of antiferromagnetic ordering in these dislocations. 
Our findings address a significant gap in the understanding of extended defects in Ge-based heterostructures and provide valuable insights for defect engineering strategies in electronic, magnetic, and spintronic device applications.

\section{METHODS}

We performed \textit{ab initio} calculations within the DFT framework as implemented in the Vienna Ab initio Simulation Package (VASP)~\cite{kresse1996efficient, kresse1996efficiency} to investigate the electronic structure of Ge containing dislocations. Projector-augmented wave (PAW) pseudopotentials were employed, treating the 4s$^2$4p$^2$ electrons as valence, with a plane-wave energy cutoff of 250 eV. Non-collinear spin-orbit coupling (SOC) was included in all calculations. To ensure a correct description of the spin–orbit splitting characteristic of Ge, calculations were also performed for primitive bulk Ge using pseudopotentials both with and without Ge 3d semi-core states, yielding no differences in the resulting spin–orbit splittings.
Structural relaxations were carried out using the revised Perdew-Burke-Ernzerhof for solids (PBEsol) exchange-correlation functional~\cite{perdew2008restoring, csonka2009assessing, zhang2018performance}. Both atomic positions and supercell vectors were optimized until reaching an electronic energy convergence threshold of $10^{-8}$ eV, guaranteeing forces per atom below 1 meV/Å in our systems. The electronic structures were computed using the meta-GGA approach of Tran and Blaha, based on the modified Becke–Johnson (mBJ) exchange potential~\cite{perdew2005prescription, tran2007band, tran2009accurate, becke2006simple}, which has been shown to provide reasonable band gaps for bulk Ge and has also been used for other polytypes~\cite{rodl2019accurate} and extended defects in Ge~\cite{Fadaly2021, Scalise2024}. Indeed, this DFT setup provides an accurate band structure at reasonable computational cost. For bulk Ge we obtained a band gap of 0.69 eV, in good agreement with the experimental low-temperature value of 0.74 eV~\cite{kittel2005introduction, ioffe_ge_database}. We note that this setup does not correctly reproduce the indirect character of the Ge gap; however, for the purposes of the present study, this limitation is not critical, as our primary focus is on the energy position of the dislocation-derived defect states relative to the bulk band gap, which is governed by the overall gap magnitude rather than by the precise dispersion of the conduction band. When necessary, the band structure was unfolded as described and implemented in the \textit{fold2Bloch} software~\cite{rubel2014unfolding, wang1998majority}. 

The dislocation cores were modeled within periodic supercells oriented along the $\left[121\right]$, $\left[1\overline{1}1\right]$, and $\left[\overline{1}01\right]$ directions, hereafter referred to as the $x$, $y$, and $z$ axes. The dislocation line is aligned along the $z$ ($\left[\overline{1}01\right]$) direction. Since periodic boundary conditions require a dipole configuration to compensate for the Burgers vector~\cite{cai2001anisotropic}, the supercell dimensions were chosen to be as large as computationally feasible to minimize the effect of the elastic interactions between dislocation cores. While fully aware of the presence of a strong elastic field in the cells, the primary concern for this study is to ensure that interactions do not significantly alter the reconstructions and the electronic properties of the dislocation cores, specifically the defect-induced energy levels within the band gap.
To address this and distinguish core-specific electronic features from spurious effects, two complementary supercell arrangements were employed. The first setup consists of supercells with identical dislocation dipoles (shuffle–shuffle or glide–glide), while the second uses a mixed shuffle–glide dipole. The identical-dipole configuration provides a model representative of strongly coupled dislocations of the same type, capturing specific electronic and magnetic effects induced by dislocation–dislocation interactions, as analyzed below. 
Conversely, the mixed dipole reduces the overlap between defect-localized wavefunctions, which would be identical and more prone to coupling in the symmetric configuration, and avoids imposing specific symmetries on the supercell. This asymmetric arrangement minimizes electronic coupling between cores, providing a more accurate model of an isolated single dislocation. The systematic comparison of results from both configurations enables the isolation of intrinsic core properties from the effects of the dipole arrangement. The comparison was also made with a supercell containing two shuffle-shuffle dislocations in a quadrupole configuration. Further information is to be found in Appendix A. To our knowledge, this represents the first comprehensive use of complementary supercell models to investigate the electronic, spin, and magnetic properties of dislocation cores at this level of detail.\\
Specifically, for the shuffle–shuffle dipole we used a supercell of $52.05 \times 29.52 \times 4.07$ Å (270 atoms), for the glide–glide dipole a $55.83 \times 39.63 \times 8.06$ Å (768 atoms) supercell, and for the mixed shuffle–glide dipole a $52.2 \times 29.7 \times 8.06$ Å (538 atoms) supercell, with the dislocation line along the $z$-axis. The Brillouin zone was sampled using $\Gamma$-centered k-point meshes of dimensions $1 \times 1 \times 4$ for the structural relaxation of the shuffle-shuffle dipole supercell, $1 \times 1 \times 2$ for the structural relaxation of the glide-glide and mixed shuffle-glide supercells and $2\times2\times12$ for the calculation of electronic properties~\cite{monkhorst1976special}.\\

\section{RESULTS}

\subsection{Core reconstructions of 60° dislocations}

60° dislocations in diamond-cubic semiconductors belong to two distinct structural families, as discussed earlier: the shuffle and glide sets. For the shuffle set, Pizzagalli et al.~\cite{pizzagalli} proposed three possible atomic reconstructions of the dislocation core in Si, labeled S1, S2, and S3. We investigated their relative thermodynamic stability in Ge through DFT structural optimizations. The differences in core energy were evaluated as half the difference in total energy between two otherwise identical supercells containing distinct dislocation dipoles, normalized by the dislocation line length. To account for the stoichiometric difference between shuffle and glide cores, which differ by a single Ge atom per dislocation period, we corrected the total energy by subtracting the bulk Ge chemical potential, $\mu_{\text{Ge}}$, from the glide configuration.

\begin{figure}
\centering
\includegraphics[width=\linewidth]{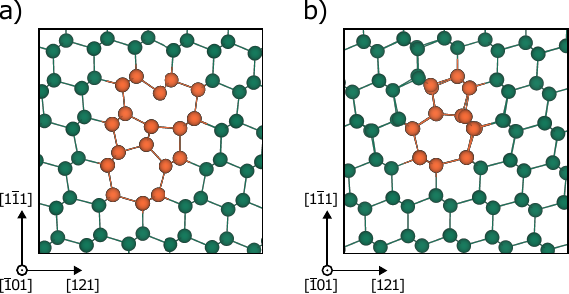}
\caption{a) S3 shuffle and b) glide core reconstructions of the 60° dislocation with a planar view normal to the dislocation line. Core atoms are highlighted in orange.}
\label{fig:core_structures}
\end{figure}

Our calculations show that the S1 reconstruction is unstable and spontaneously transforms into S2 upon ionic relaxation, while S2 remains stable as a local energy minimum. However, S2 lies 0.12 eV/Å higher in energy than S3 (Fig.~\ref{fig:core_structures}a), establishing S3 as the most favorable shuffle reconstruction. Comparing the two families, the glide reconstruction (Fig.~\ref{fig:core_structures}b) is significantly more stable than S3, with an energy difference of 1.22 eV/Å in favor of the glide configuration. 

We recall that shuffle dislocations can undergo a transformation to the glide set through interaction with vacancies, which passivate dangling bonds and induce a double-periodicity reconstruction along the dislocation line~\cite{hornstra1958dislocations, pizzagalli,hull2011introduction,barbisan2022atomic}. This mechanism provides a kinetic pathway from shuffle to glide configurations. Although the glide configuration is energetically more stable, we also consider the most stable shuffle-set reconstruction, S3, which may be kinetically hindered from transforming into the glide configuration by the lack of vacancies. In fact, S3 shuffle and glide reconstructions emerge as the primary candidates of the 60° dislocation family to be present in epitaxial Ge layers, with the potential to significantly impact their electronic and optical properties. 

\subsection{Electronic properties and spin–orbit coupling in S3 and glide core dislocations}

\begin{figure*}
\centering
\includegraphics[width=1\linewidth]{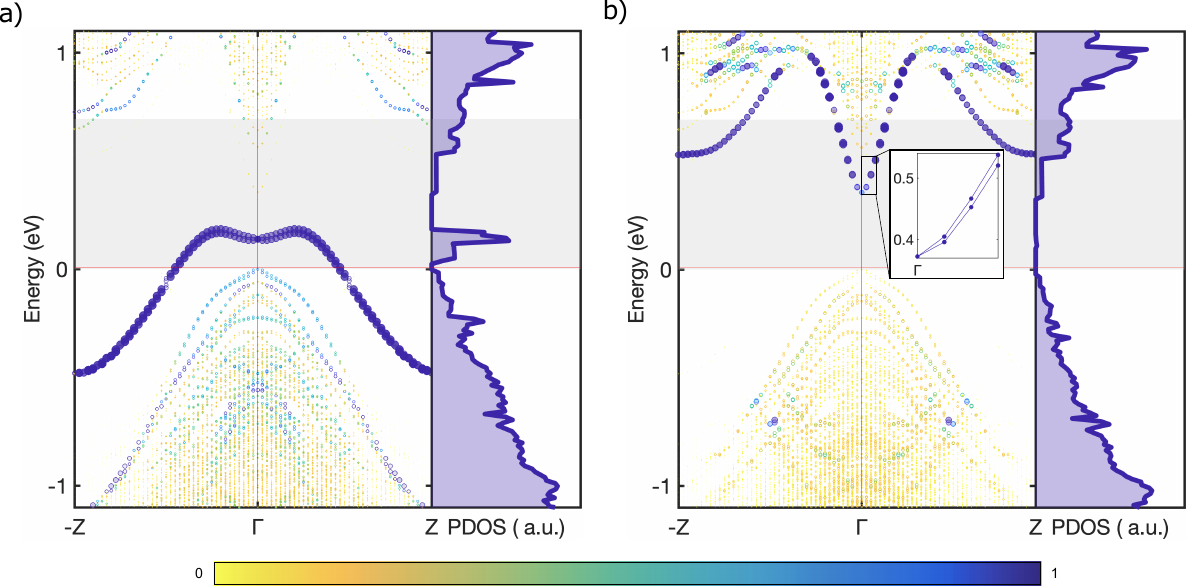}
\caption{Band structure and partial density of states (PDOS) of the defective supercell, projected onto the dislocation-core atoms. The band structures are plotted along the $-Z$ -- $\Gamma$ -- $Z$ path, $Z$ being a high-symmetry point in the reciprocal space of the supercell, with relative coordinates $\pm Z = (0,0,\pm0.5)$. The valence band maximum is set to 0 eV, and the solid red line indicates the Fermi level. Panels (a) and (b) show the unfolded band structures projected onto the shuffle S3 dislocation core atoms and glide dislocation core atoms, respectively. The color scale represents the combined weight of the band unfolding and the projection onto the dislocation core atoms, obtained through the normalized local density of states. The gray shaded region indicates the bulk Ge bandgap, with the VBM set to 0 eV. In panel (b), the inset shows a magnified view of the region delimited by the black box, where the energy splitting of the glide dislocation bands has been enhanced by a factor of three to improve visibility.
}
\label{fig:band_mix}
\end{figure*}

We now discuss the electronic properties of the S3 and glide cores. Figure~\ref{fig:band_mix} presents the calculated partial density of states (PDOS) and electronic band structures for both reconstructions, obtained from the mixed-dipole supercell. Since the dislocation is intrinsically periodic only along its line direction, the physically relevant dispersion is the one-dimensional dispersion along this direction. Accordingly, both the S3 and glide core states exhibit characteristic dispersion along the $\Gamma$–$X$ direction of the Ge Brillouin zone, which corresponds to the dislocation-line orientation. In the directions transverse to the line, the defect behaves as an isolated point-like core, and the corresponding defect states remain essentially flat within the periodic-supercell representation. The high-symmetry point of the supercell along the line direction, labeled $Z$, is located approximately halfway along the $\Gamma$–$X$ path.
A technical consideration arises from the structural difference between the two cores: as discussed earlier, the glide reconstruction exhibits a doubled structural periodicity along the dislocation line compared with S3. To accommodate both core structures simultaneously, the mixed-dipole supercell adopts a doubled periodicity along the line direction, specifically with two primitive periods of S3 matching one period of the glide core. This doubling causes band folding along the $\Gamma$–$Z$ path of the supercell reciprocal space. To facilitate interpretation, we performed an unfolding procedure to generate the bands shown in Fig.~\ref{fig:band_mix}.

\begin{figure*}
\centering
\includegraphics[width=1\linewidth]{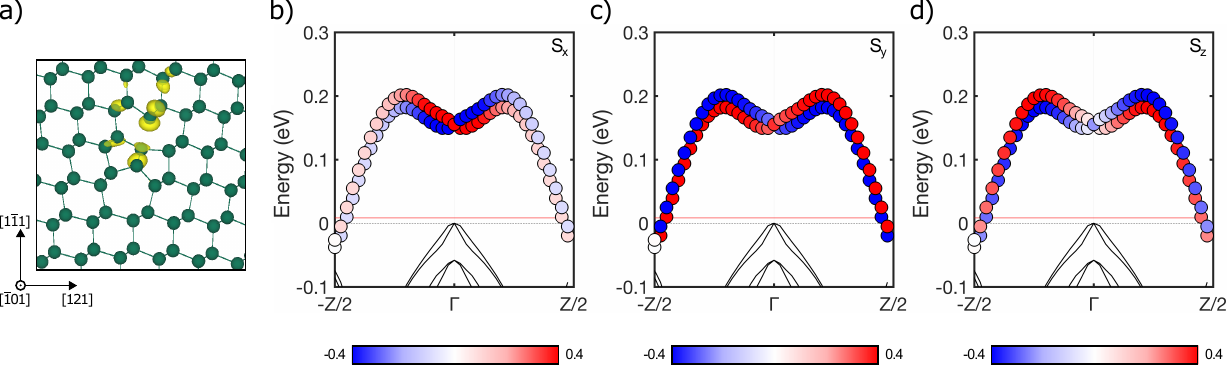}
\caption{a) Charge density (yellow surfaces) localization on the shuffle S3 core; b), c) and d) Spin-resolved band structure of the mixed dislocation dipole, in the energy range around the shuffle S3 defects states. Black solid lines denote bulk states, while dotted lines correspond to the shuffle S3 states. The valence band maximum is set to 0 eV, and the solid red line indicates the Fermi level. The color scale represents the normalized expectation values of the spin projections $S_x$ (panel b, with $x$ along $\left[121\right]$ direction), $S_y$ (panel c, with $y$ along $\left[1\overline{1}1\right]$ direction) and $S_z$ (panel d), with z parallel to the dislocation line ($\left[\overline{1}01\right]$ direction).}
\label{fig:rashba_split}
\end{figure*}

By projecting the band structure onto the core atoms of the S3 dislocation, the electronic structure reveals highly localized defect states well separated from the bulk bands and positioned deep within the Ge band gap (Fig.~\ref{fig:band_mix}a). These states manifest as a pair of dispersive bands around the $\Gamma$ point, with a dispersion spanning an energy range of 0.66 eV, particularly pronounced away from $\Gamma$ where they extend below the valence-band maximum (VBM). 
Remarkably, at the $\Gamma$ point these bands are degenerate, as shown in the spin-resolved band structure plotted in Fig. \ref{fig:rashba_split}. Conversely, they split along the $\Gamma$ -- $Z$ path, with a maximum splitting of 20 meV, and show opposite spin polarization, evidence of spin-orbit coupling. The spin texture analysis presented in the Appendix C reveals a striking helical spin pattern consistent with a Rashba-Dresselhaus spin-orbit coupling (SOC). This helical texture, with spins rotating coherently around the dislocation line, is likely to originate from the screw component of the 60° dislocation, as it closely resembles the behavior reported by Liu et al.~\cite{hu2018ubiquitous} for pure screw dislocations in Ge. The presence of such pronounced SOC in a mixed-character 60° dislocation demonstrates that the screw component dominates the spin physics at the shuffle core, also in the presence of an edge component.

The glide reconstruction also manifests electronic states localized at the dislocation core. However their proximity to the conduction band of the defected supercell, particularly close to the $\Gamma$ -- $Z$ middle point, significantly affects their characterization (Fig.~\ref{fig:band_mix}b). These states appear as a pair of dispersive bands just below, and often overlapping with, the lowest-energy branches of the conduction band, with a dispersion extending over approximately 0.6 eV. In this case, the energetic splitting between the two bands due to SOC is very small, reaching a maximum of 7 meV, which makes them nearly indistinguishable within the resolution of the $Z$--$\Gamma$--$Z$ band-structure plot of Fig.~\ref{fig:band_mix}b. This splitting is nevertheless visible in the inset, where a zoom around the $\Gamma$ point clearly shows a degeneracy at $\Gamma$ and an increasing energy splitting along the k path (multiplied by a factor of three to enhance visibility). This behavior suggests Rashba-Dresselhaus spin–orbit coupling (SOC) similar to that observed in the S3 shuffle core and in screw dislocations~\cite{hu2018ubiquitous}. However, the strong overlap of these states with the conduction band prevents a clear resolution of the spin texture, making it far less pronounced and more difficult to isolate compared to the well-separated deep-gap states of the S3 reconstruction.

\subsection{Magnetic ordering in shuffle S3 dislocations}
The shuffle S3 dislocation core hosts a state that is spatially localized in the plane perpendicular to the dislocation line, as evidenced by the charge density analysis shown in Fig. \ref{fig:rashba_split}a, which reveals strong confinement at the core atoms. Further confirmation of this confinement is given by the Inverse Participation Ratio analysis (see Appendix B). At the same time, the periodic repetition of this locally confined orbital along the dislocation line gives rise, through a tight-binding-like hopping between adjacent core sites, to a dispersive band visible in Fig.~\ref{fig:band_mix}a. This band dispersion is therefore a consequence of the one-dimensional periodicity of the dislocation rather than of intrinsic delocalization of the individual core orbital. A sharp peak in the density of states appears at approximately 0.15 eV above the VBM, and the Fermi level lies slightly above the VBM, resulting in partial filling of the defect band. 
The combination of a large defect-related peak in the density of states, partial band filling, and strong spatial localization provides favorable conditions for the onset of magnetic ordering. Within a simple Stoner model, when localized states cross the Fermi level with a sufficiently high density of states, the strong spatial confinement enhances the exchange interaction, and the system can spontaneously develop spin polarization: the total energy is lowered by creating an imbalance between spin-up and spin-down occupation, resulting in an exchange splitting between majority and minority spin bands~\cite{zhang2013intrinsic}. The energy gain from the exchange interaction compensates for the kinetic energy cost of promoting electrons to higher-energy states, stabilizing the magnetic configuration. Similar behavior has been observed in other one-dimensional systems where localized edge or defect states at the Fermi level induce magnetic ordering, including grain boundaries in two-dimensional metal dichalcogenides~\cite{zhang2013intrinsic} and zigzag edges in bismuth nanoribbons~\cite{naumov2023one}.

\begin{figure}
\centering
\includegraphics[width=0.8\linewidth]{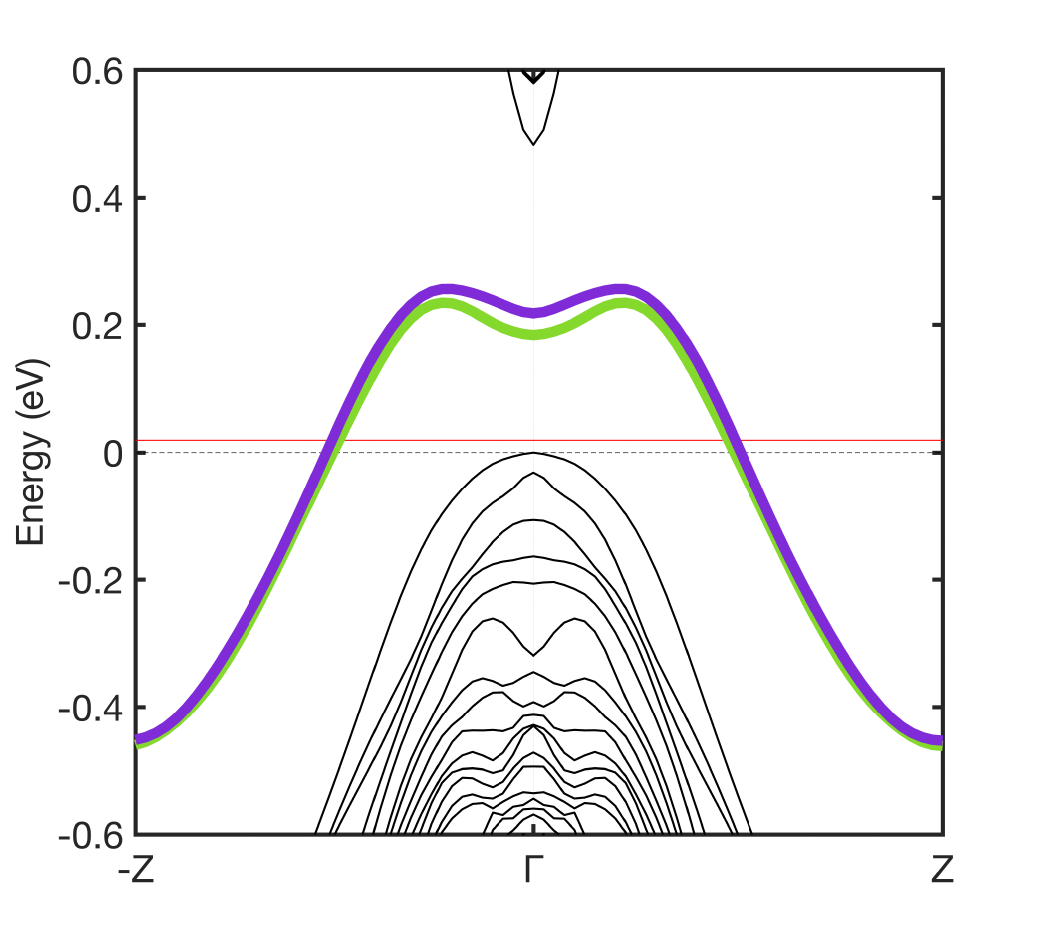}
\caption{Bandstructure of the non-magnetic 60° shuffle S3 dislocation dipole. The valence band maximum is set to 0 eV, and the solid red line indicates the Fermi level. The green and purple solid lines denote two pairs of degenerate bands associated with the dislocation cores.}
\label{fig:nonmagn}
\end{figure}

To assess the magnetic coupling between two 60° shuffle S3 dislocations, we investigated supercells containing a dipole composed of two identical shuffle dislocations. The electronic states associated with these dislocations, obtained from both the mixed-dipole (Fig.~\ref{fig:band_mix}a) and shuffle-shuffle dipole supercells (Fig.~\ref{fig:nonmagn}), are essentially identical in terms of dispersion, shape, and energetic position within the Ge band gap. This excellent agreement highlights the robustness of the calculated defect states with respect to the choice of supercell model. 

Our calculations yield two self-consistent solutions, non-magnetic (NM) and antiferromagnetic (AFM), whose energies differ by only about 1 meV per core atom. This energy difference is comparable to the numerical convergence threshold, so the two solutions should be regarded as nearly degenerate within the accuracy of the present calculations. Nevertheless, the AFM configuration reveals a richer physical picture through the interplay of exchange and spin-orbit coupling. In both cases, four mid-gap bands are observed rather than the two observed in the mixed-dipole configuration, as a direct consequence of the presence of two identical dislocations in the supercell. Notably, no analogous $IT$-protected degeneracy is associated with a ferromagnetic alignment of the two cores. Consistently, calculations initialized with ferromagnetic moments always relaxed to the non-magnetic solution, and no distinct self-consistent ferromagnetic state was obtained.

In the non-magnetic configuration (Fig.~\ref{fig:nonmagn}), the four bands appear pairwise degenerate. Each pair corresponds to the spin-up and spin-down states associated with the same dislocation core, a degeneracy enforced by time-reversal symmetry $T$, which is preserved in the system. A small splitting between the two pairs (0.03 eV) arises from a subtle breaking of spatial inversion symmetry $I$. 

\begin{figure*}
\centering
\includegraphics[width=1\linewidth]{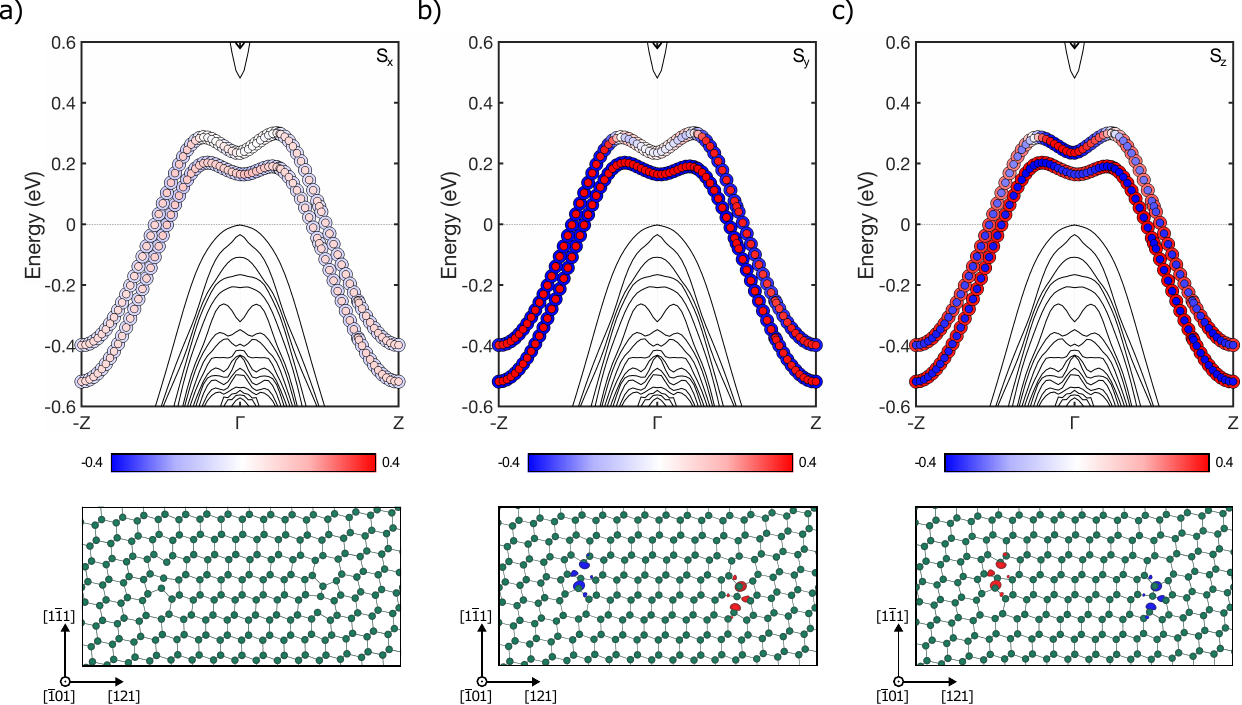}
\caption{Spin resolved band structures of the antiferromagnetic 60° shuffle S3 dislocation dipole. Dislocation bands are shown as circles, while the bulk states are represented by the black curves. The valence band maximum is set to 0 eV. The Fermi level lies 0.33 meV above it and is thus indistinguishable from the valence band maximum within the energy resolution of the plot. Color code represents the expectation values of the spin projections for the $S_x$ spin projections (panel a, with $x$ along $\left[121\right]$ direction), $S_y$ spin projection (panel b, with $y$ along $\left[1\overline{1}1\right]$ direction) and $S_z$ spin projection (panel c), with z parallel to the dislocation line ($\left[\overline{1}01\right]$ direction. The corresponding (directional) spin densities are shown below. Blue (red) color indicates the negative (positive) isovalues in the isosurface plots.}
\label{fig:afm}
\end{figure*}

This symmetry breaking originates from a numerical artifact inherent to the finite supercell: although the two dislocation cores are nominally identical by construction, the elastic strain fields they generate distort the supercell in such a way that, upon structural relaxation, the two cores become slightly nonequivalent. This effect is minimal, comparable to numerical noise and not discernible from the relaxed atomic structure, yet sufficient to lift the degeneracy between the electronic states localized on the two dislocations. Despite this small splitting, the cores remain nearly equivalent, so the defect states form bonding- and antibonding-like combinations that are delocalized over both dislocations. The energy splitting induced by the residual inversion asymmetry is much smaller than the electronic coupling between the cores, and consequently the charge density remains substantially shared across both defects.

In the antiferromagnetic configuration (Fig.~\ref{fig:afm}), time-reversal symmetry $T$ is broken, lifting the spin degeneracy within each individual dislocation core through exchange interaction. This exchange-induced spin splitting is substantial, reaching $\Delta E_{\text{ex}} = 0.07$ eV at the $\Gamma$ point, and exhibits k-dependent behavior along the $\Gamma$--$Z$ path, reflecting the electronic structure of the localized defect states. 

Remarkably, while $T$ and $I$ are individually broken, the combined symmetry $IT$ remains preserved in the antiferromagnetic arrangement~\cite{naumov2023one}. This combined symmetry enforces a twofold degeneracy at each k-point, effectively restoring the degeneracy between states localized on the two different dislocation cores.
In the antiferromagnetic state, this degeneracy now involves states of opposite spin on opposite cores, satisfying the relation $E(k, \uparrow, 1) = E(k, \downarrow, 2)$, where indices 1 and 2 label the two dislocations forming the dipole. Consequently, the small 0.03 eV splitting observed between core states in the non-magnetic case vanishes in the antiferromagnetic configuration, replaced by the larger 0.07 eV exchange splitting between opposite spins on the same core. This symmetry-protected degeneracy in the antiferromagnetic state may confer enhanced robustness to the electronic structure of the dislocation cores, potentially stabilizing the defect levels against small perturbations and local disorder.

The magnetization is predominantly aligned along $y$ and $z$, $\left[1\overline{1}1\right]$ and $\left[\overline{1}01\right]$ crystallographic directions, respectively. Conversely, the $x$~-~component ($\left[121\right]$ direction) is negligible,  as clearly shown in the spin density plots of Fig.~\ref{fig:afm}. The total magnetic moment $\mathbf{M}$ at each dislocation core is approximately $0.69\,\mu_{\text{B}}$, indicating a substantial local spin polarization, despite the relatively small energy difference between the non-magnetic and antiferromagnetic states.
The emergence of significant local magnetic moments at the shuffle dislocation cores, combined with the characteristic Rashba–Dresselhaus spin-orbit coupling, demonstrates that dislocations in Ge can impart qualitatively new physical properties to the material rather than simply degrading its performance. The topological stability of these extended defects, coupled with their tunable electronic and magnetic properties via Fermi level modulation or the application of external electromagnetic fields, makes them particularly attractive as building blocks for spintronic functionalities in Ge-based heterostructures.

\subsection{Deep defect states induced by dislocation cores}

\begin{figure}
\centering
\includegraphics[width=1\linewidth]{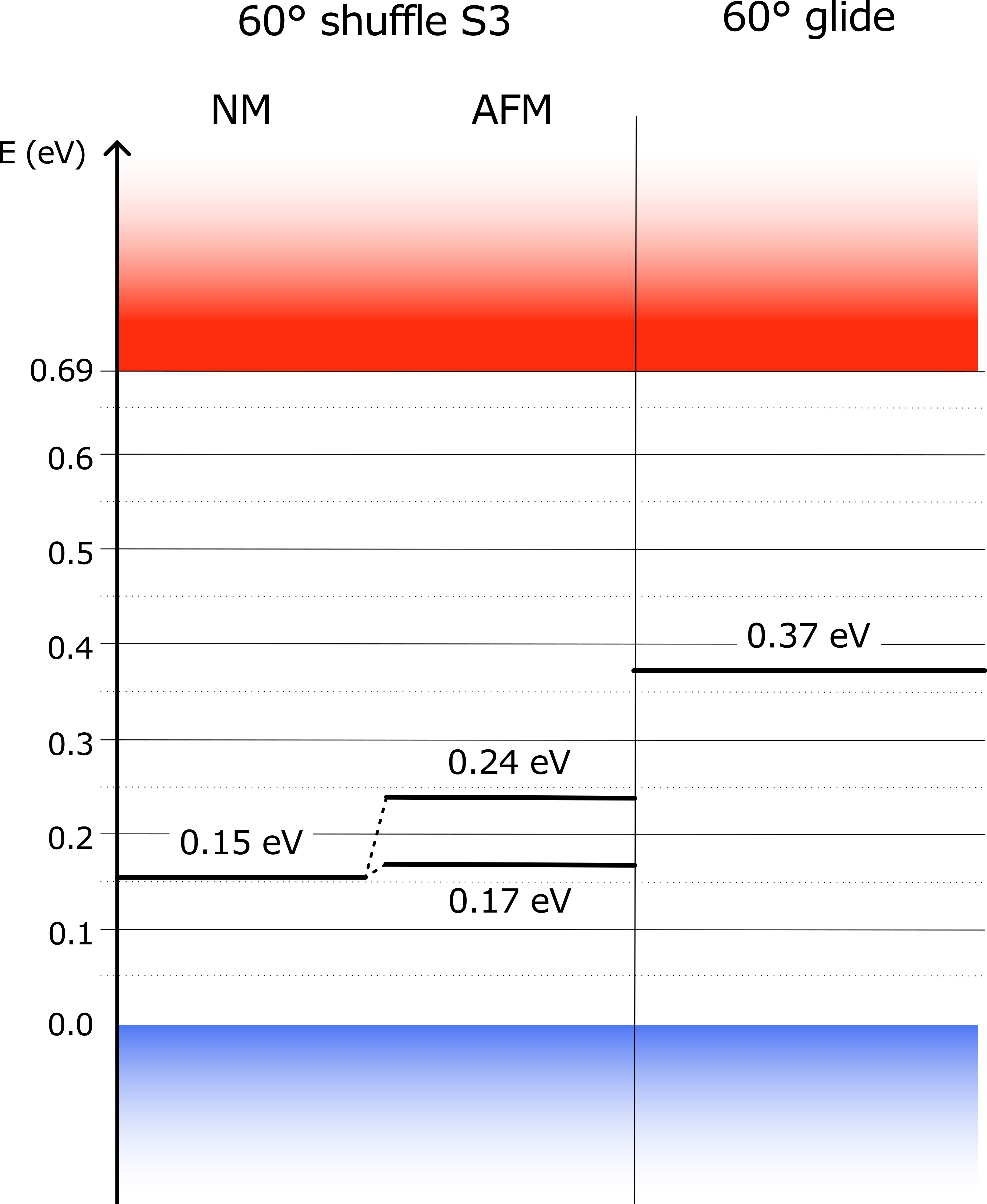}
\caption{The graph summarizes the defect levels at the $\Gamma$ point for the shuffle S3 reconstruction, calculated in both the NM and AFM configurations using the shuffle–shuffle dipole, and for the glide reconstruction, calculated using the glide–glide dipole. The band gap is that of bulk Ge, computed with the same computational setup. For the glide reconstruction, the reported level is consistent with the corresponding level obtained from the mixed shuffle–glide dipole.}
\label{fig:final_scheme}
\end{figure}

The band-structure calculations presented in the previous sections allow us to directly identify and characterize the electronic states associated with the dislocation cores in the band-gap region. Figure~\ref{fig:final_scheme} summarizes the positions of these states with respect to the bulk valence-band maximum (VBM), evaluated at the $\Gamma$ point, for both the 60\textdegree{} shuffle S3 and the 60\textdegree{} glide reconstructions. The reported energies correspond to the Kohn--Sham eigenvalues evaluated at $\Gamma$, thereby providing a single reference energy for each defect state.
While the band-structure plots explicitly show the dispersion along the dislocation line, here we report a single value. This choice reflects the intrinsically one-dimensional character of an isolated dislocation: the defect states are localized around the core in the directions transverse to the dislocation line, whereas the physically meaningful band dispersion is associated with the periodic direction along the line. We therefore use the $\Gamma$-point eigenvalue as a well-defined single-value descriptor of each defect state, enabling a direct comparison between the two reconstructions and providing a consistent reference for comparison with experimentally reported defect or trap energies.

The shuffle S3 reconstruction introduces in-gap states whose character depends on the magnetic configuration. In the non-magnetic solution, a single defect level appears at approximately $0.15$ eV above the VBM. When an antiferromagnetic ordering is imposed on the dislocation core, this level splits into two distinct states at $0.17$ and $0.24$ eV above the VBM, reflecting the symmetry breaking introduced by the spin ordering. In both the non-magnetic and antiferromagnetic solutions, this defect band is crossed by the Fermi level along the dispersive direction and is therefore half-filled (metallic) in both configurations.
The glide reconstruction, consistent with its more dispersive character and smaller PDOS within the gap, shows a single level at approximately $0.37$ eV above the VBM at the $\Gamma$ point, while the remaining dispersive state is resonant with or very close to the CBM. Unlike the S3 case, the glide defect band lies entirely above the Fermi level along the dispersive direction and is therefore unoccupied.

These defect levels can be directly compared with experimentally measured trap energies in Ge ~\cite{arguirov2014luminescence, kittler2011photoluminescence, pezzoli2014ge, tanaka1996photoluminescence, grillot1996effect, hubbard1980electrical, claeys2009extended}. 
Table~\ref{tab:comparison} summarizes the calculated levels alongside selected experimental values from the literature obtained by optical absorption spectroscopy, DLTS and Hall-effect measurements on deformed or implanted Ge samples~\cite{arguirov2014luminescence, kittler2011photoluminescence, pezzoli2014ge, tanaka1996photoluminescence, grillot1996effect, hubbard1980electrical, claeys2009extended}. A direct one-to-one quantitative comparison remains inherently challenging: epitaxial Ge layers typically contain multiple types of dislocations, point defects, and impurities whose contributions are difficult to disentangle, and the inherent accuracy of DFT-based calculations for absolute gap positions is limited. Nevertheless, the systematic agreement between the calculated levels and the experimentally observed traps, with deviations within 20~meV, suggests that these experimental features may tentatively be attributed to the intrinsic electronic activity of 60° dislocation cores: the calculated level at $0.15$~eV (S3, NM) is consistent with the shallow dislocation-related acceptor band observed near $E_V + 0.13$~eV by optical absorption in Ge~\cite{claeys2009extended}. The two AFM levels at $0.17$ and $0.24$~eV agree well with the DLTS hole traps H3 ($E_V + 0.17$~eV) and H1 ($E_V + 0.23$~eV)~\cite{claeys2009extended}, commonly observed in p-type Ge. The level at $0.25$~eV is in good agreement with the Hall-effect-derived acceptor band near $E_V + 0.25$~eV~\cite{claeys2009extended}. The glide level at $0.37$~eV is in close agreement with the DLTS trap H2 at $E_V + 0.37$~eV \cite{claeys2009extended}, while an additional point-defect-like feature at $E_V + 0.39$~eV is identified in deformed Ge \cite{claeys2009extended}. Note that, since all the compared levels are referenced to the VBM, the comparison is less sensitive to the slight underestimation of the experimental band gap.

\begin{table}[h]
\caption{Comparison of the calculated defect levels shown in Fig.~\ref{fig:final_scheme} with experimental trap energies reported for dislocations in Ge \cite{claeys2009extended}. All energies are referenced to the VBM ($E_V$). The absolute difference between the experimental value and the calculated energy level is denoted by $|\Delta|$.  }
\label{tab:comparison}
\begin{tabular}{lcc}
\hline
Exp.(eV) & $|\Delta|$ (meV) & Method \\
\hline
0.13 & 20 & Optical absorption\\
0.17 & 0  & DLTS, p-Ge \\
0.23 & 10 & DLTS, p-Ge \\
0.25 & 20 & Hall-effect \\
0.37 & 0  & DLTS, p-Ge \\
0.39 & 20 & DLTS, deformed Ge\\
\hline
\end{tabular}
\end{table}

\section{CONCLUSIONS}

In summary, we present a comprehensive first-principles investigation of the structural stability and electronic properties of 60° perfect dislocations in germanium, considering both glide-set and shuffle-set reconstructions. Our DFT calculations using the mBJ meta-GGA functional reveal that the electronic properties of these two reconstructions differ markedly. The glide reconstruction exhibits a nearly bulk-like electronic character with dislocation-induced states that overlap strongly with the valence band, resulting in minor perturbation of the Ge band structure. In contrast, the shuffle S3 reconstruction introduces highly localized dispersive bands within the Ge bandgap, positioned approximately 0.15 eV above the VBM. These localized states cross the Fermi level and exhibit substantial spin-orbit coupling effects, displaying characteristic Rashba-Dresselhaus helical spin textures analogous to those reported for screw dislocations. 

The combination of a large defect-related peak in the density of states lying close to the Fermi level and strong spatial localization of the defect states provides favorable conditions for magnetic ordering through a Stoner-like mechanism. In the shuffle-shuffle dipole configuration, we find that the system can stabilize an antiferromagnetic state with local magnetic moments of approximately $0.69\,\mu_{\text{B}}$ per dislocation core and an exchange splitting of $\Delta E_{\text{ex}} = 0.07$ eV between opposite spin states. Notably, the antiferromagnetic configuration restores the exact degeneracy between electronic states localized on the two dislocation cores through combined inversion-time-reversal ($IT$) symmetry, a degeneracy that was weakly broken in the non-magnetic state due to residual strain-induced asymmetry in the finite supercell. This symmetry-enforced degeneracy suggests that the electronic states of coupled identical dislocation cores may exhibit enhanced robustness against local perturbations and structural disorder, potentially providing intrinsic stability to the spin-polarized defect levels against environmental fluctuations. 

These results demonstrate that dislocations in Ge can impart qualitatively new physical properties, such as intrinsic magnetism, spin-orbit-coupled electronic states, and topological spin textures, rather than simply degrading material performance. The topological stability of these extended defects, coupled with their tunable electronic and magnetic properties, positions them as promising building blocks for spintronic functionalities in Ge-based heterostructures.

These findings highlight the crucial role of core reconstructions in dictating the functional properties of dislocations in germanium. Our results address a significant gap in understanding extended defects in Ge and demonstrate that careful control of dislocation core structure can enable novel quantum and spintronic functionalities in this technologically relevant semiconductor platform.

\section*{Acknowledgements}

The authors acknowledge the CINECA consortium under the ISCRA initiative for the availability of high-performance computing resources and support. 

F.M. acknowledges financial support by the European Union – NextGenerationEU – Investment 1.1, M4C2 - Project Title “NANOSEES” n. P2022YM8J3 – CUP D53D23018720001 - Grant Assignment Decree No. 1389 adopted on 1st September 2023 by the Italian Ministry of University and Research (MUR).

\section*{Data Availability}
The data that support the findings of this article are openly available \cite{regazzoni_2026_22796437}.

\appendix

\section{Electronic characteristics of different structural configurations}

The influence of elastic interactions between the two dislocations of the dipole on the electronic structure should in principle be assessed through a systematic convergence study as a function of supercell size. However, the simulation cells considered in this work are already at the limit of current computational capabilities, making such a study unfeasible. We therefore adopted a different approach, comparing electronic properties across different dipole arrangements that are characterized by distinct relative dislocation positions and elastic coupling.

\begin{figure}[h]
\centering
\includegraphics[width=1\linewidth]{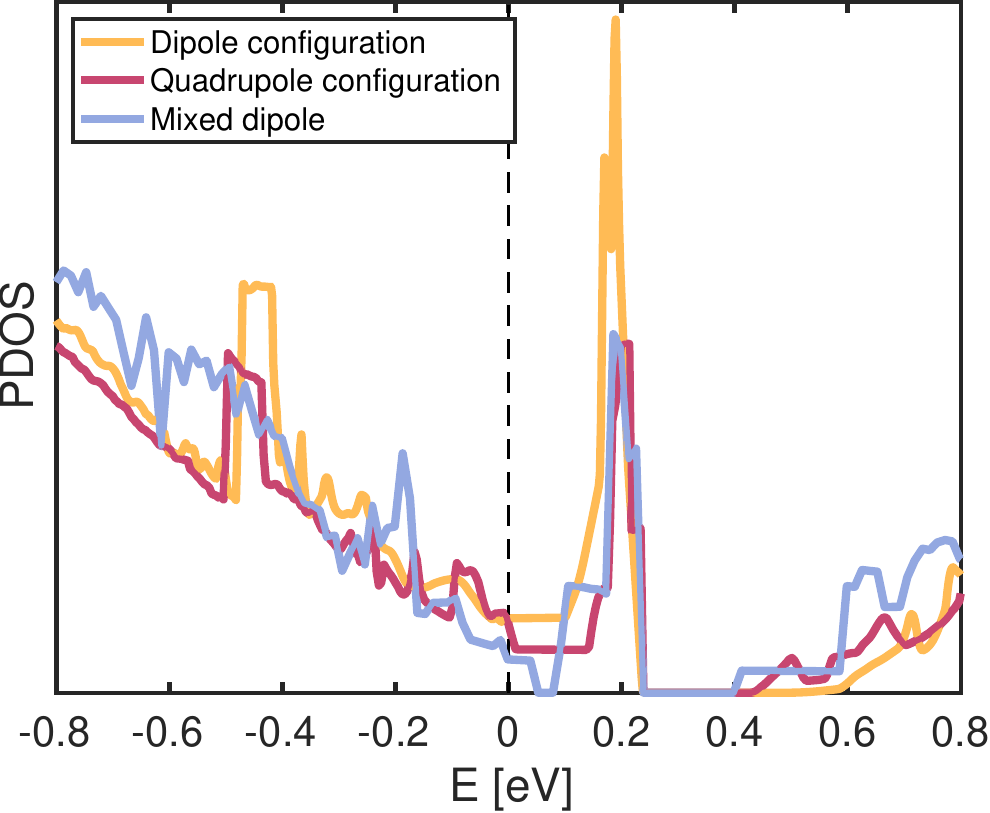}
\caption{PDOS projected onto the atoms of the 60° shuffle S3 dislocation cores for different supercell geometries, i.e. the S3-S3 dipole, the S3-S3 quadrupole and the mixed S3-glide dipole.}
\label{fig:S1}
\end{figure}

Specifically, we considered three configurations: the S3–S3 dipole, the S3–S3 quadrupole, and the S3-glide dipole (labeled "mixed" in the following). The comparison between the S3–S3 dipole and quadrupole allows us to isolate the effect of elastic self-interaction, since both configurations involve the same defect type but with different mutual positions and therefore different residual elastic fields. As shown in Fig.~\ref{fig:S1}, the PDOS of these two configurations is in close agreement, particularly for the in-gap states, demonstrating that the residual elastic interaction has a negligible effect on the electronic properties of interest.

The comparison between the S3–S3 dipole and the S3-glide mixed dipole serves a complementary purpose: it demonstrates that the PDOS projected onto the S3 dislocation core is not affected by the presence of a different dislocation type in the same supercell. This confirms that the contributions of the two defects to the in-gap states can be successfully decoupled, which is essential for the interpretation of the results presented in this manuscript.

\section{Inverse Participation Ratio analysis}
\begin{figure}[h]
\centering
\includegraphics[width=1\linewidth]{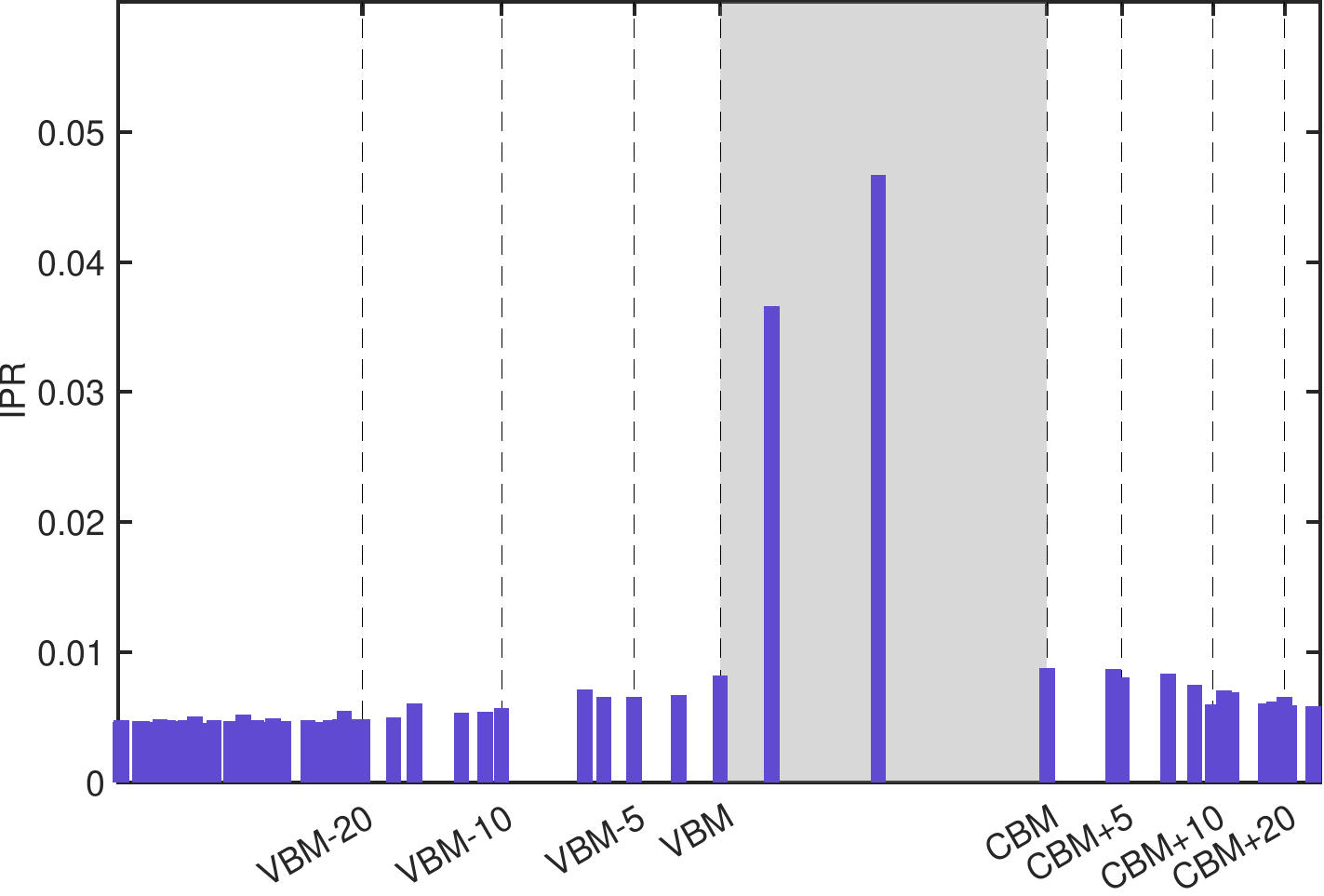}
\caption{Inverse participation ratio (IPR) for the S3–S3 dislocation dipole in the antiferromagnetic configuration, resolved as a function of band index relative to the VBM and CBM}
\label{fig:S4}
\end{figure}

The IPR is a measure of the inverse of the number of sites involved in the state with energy E, commonly used in the study of defective systems to evaluate the degree of localization of their electronic levels \cite{wilhelmer2023over, atta2004systematic, ogawa2011ab}. It is defined as
\begin{equation*}
    IPR_{nk}=\frac{\sum_a |c_{nka}|^4}{(\sum_a|c_{nka}|^2)^2}
\end{equation*}
with $n$ band index, $k$ kpoint index, $a$ atom index and $c_{nks}$ coefficient of the wavefunction. The IPR can assume values between $1/N$ and 1, where $IPR=1$ means a perfectly localized state and $IPR = 1/N_{tot}$ means a perfectly delocalized state.
To investigate the localization and properties of the electronic states associated with the 60° shuffle S3 dislocation, we compute the inverse participation ratio (IPR) for the S3–S3 dipole supercell in the antiferromagnetic configuration. Figure \ref{fig:S4} reports the IPR as a function of the band index, referenced to the valence band maximum (VBM) and conduction band minimum (CBM). 
The two defect-associated bands are clearly characterized by IPR values significantly larger than those of all other states, consistent with their spatial localization around the dislocation core. In contrast, bulk states exhibit very small IPR values ($\sim$ 0.005), indicating full delocalization over the entire supercell. The defect states display larger IPR values, approximately 0.045. This corresponds to a spatial localization on almost 22 atoms, consistent with the dimensions of the core of the dislocations. To further reinforce the localization of the defect, a combined analysis between IPR and the charge density localization associated with the first dislocation-associated state shown in Fig.\ref{fig:rashba_split}a underlines the optimal spatial localization of the state around the dislocation core in the plane perpendicular to the dislocation line. 

\section{Spin textures in k-space}
\begin{figure*}
\centering
\includegraphics[width=0.9\linewidth]{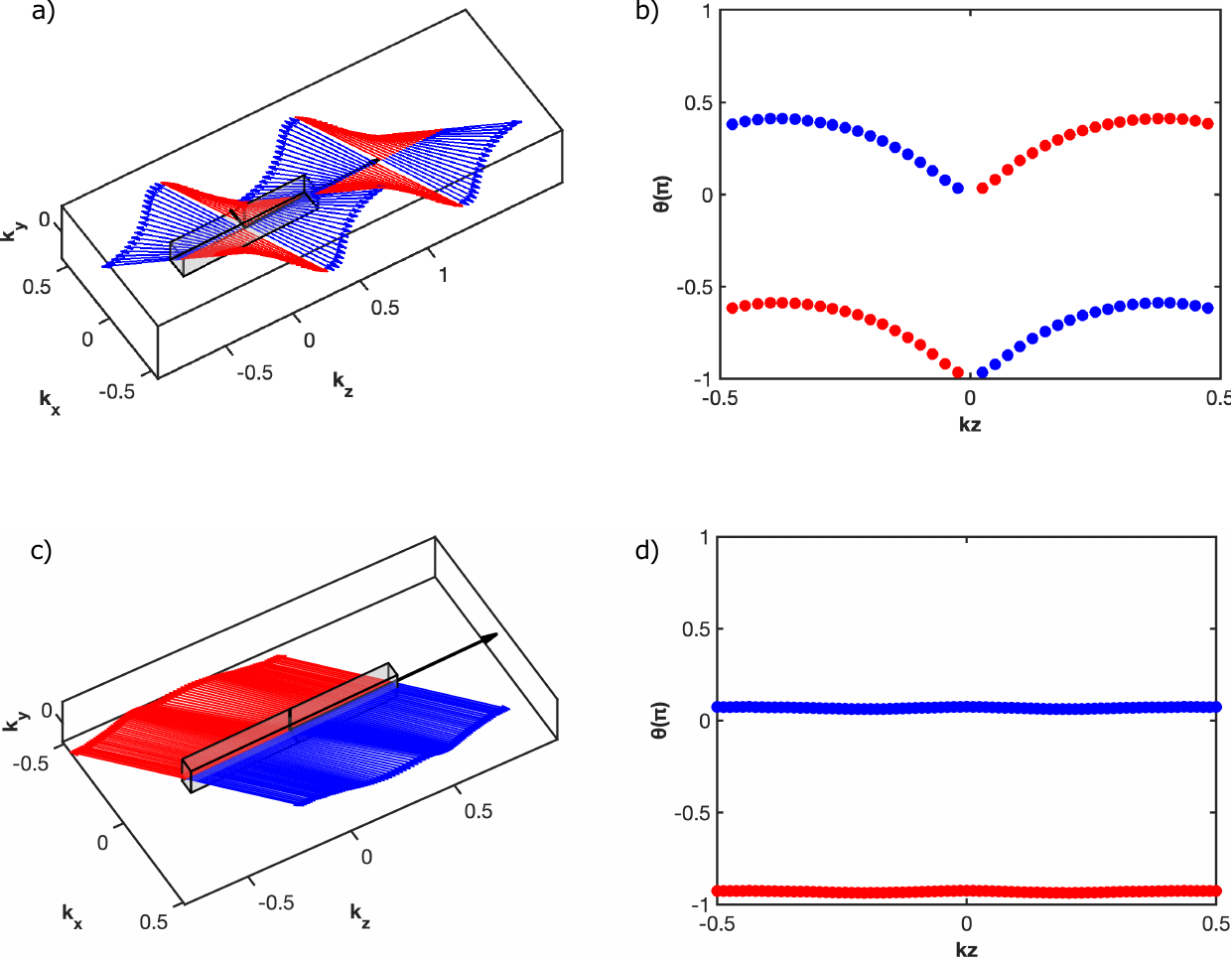}
\caption{Spin texture associated with the dislocation bands of the 60\textdegree{} shuffle S3 configuration. The bands are highlighted in red and blue. $k_z$, along the direction of periodicity of the dislocation, is the only physical crystal-momentum variable. The gray surface indicates the boundary of the Brillouin zone. $k_x$ and $k_y$ axes define a fixed geometric reference frame in which the spin vector $(S_x, S_y)$ associated with each band is represented; they do not denote independent crystal-momentum variables. Panel (a) shows the spin texture for the non-magnetic configuration, where a helical spin structure can be observed. Panel (c) shows the spin texture for the antiferromagnetic configuration, where the spins are aligned in an antiparallel arrangement. Panels (b) and (d) show the angle $\theta$ between the spin vector and the $k_x$ reference axis, as a function of $k_z$.}
\label{fig:S2}
\end{figure*}

We present and compare the spin textures associated with the dislocation-induced electronic states in the non-magnetic and antiferromagnetic (AFM) configurations, shown in Fig.~\ref{fig:S2}. Although the Brillouin zone of the supercell is three-dimensional, the dislocation is periodic only along $k_z$, so that $k_z$ is the only crystal-momentum direction along which the bands disperse. The spin orientation at each $k_z$ is therefore analyzed in terms of the angle $\theta$ between the spin vector $(S_x, S_y)$ and a fixed $x$-axis reference direction, in analogy with the customary depiction of helical or antiparallel spin textures \cite{hu2018ubiquitous}.

In the non-magnetic case (Fig. \ref{fig:S2}a), the spin texture exhibits a pronounced helical character in reciprocal space, consistent with a screw-dislocation-induced spin–orbit coupling mechanism \cite{hu2018ubiquitous}. However, the observed helical pattern deviates from an ideal periodic and perfectly symmetric winding expected for an ideal screw-symmetric system. These deviations can be attributed to several concurrent effects. First, a slight tilting of the supercell axis with respect to the dislocation line, introduced during structural relaxation, breaks the exact alignment of the effective symmetry axis. Second, the 60° dislocation geometry does not support a strict twofold screw symmetry; at most, a local sixfold rotational character can be identified, which relaxes the symmetry constraints underlying an ideal helical spin winding. Finally, additional symmetry-breaking contributions in the effective Hamiltonian, which would vanish for an isolated ideal screw dislocation, further perturb the spin structure and lead to distortions of the Rashba-like helical texture. These effects are also reflected in the $\theta(k_z)$ dependence (Fig. \ref{fig:S2}b). In contrast to the linear behavior expected for a pure screw dislocation in Ge, the mixed-character dislocation exhibits a monotonically increasing but non-linear variation of $\theta$ as a function of $k_z$, indicating a deviation from an ideal helical spin rotation.

In contrast, for the AFM configuration (Fig. \ref{fig:S2}c), an additional effective inversion symmetry emerges at the level of the full supercell, relating the two dislocations. This higher-order symmetry constrains the spin degrees of freedom and enforces an antiparallel alignment of the magnetic moments associated with the dislocations. As a result, the spin texture is no longer governed by the local screw symmetry of individual dislocations, but by the global symmetry of the supercell, leading to an antiferromagnetic arrangement rather than a helical winding. In this regime, the interplay between magnetic ordering and crystal symmetry induces an effective suppression of spin precession, resulting in a strongly constrained spin response compared to the non-magnetic case. The dislocation-induced spin–orbit coupling field therefore does not generate a rotating texture, but instead stabilizes an almost uniform spin orientation. Overall, Ge in the AFM phase exhibits a qualitatively distinct spin response with respect to the non-magnetic case, characterized by suppressed spin rotation and enhanced spin alignment along a preferred direction. This highlights the key role of magnetic ordering in controlling the symmetry and robustness of dislocation-induced spin textures, with potential implications for spin transport in defect-engineered systems.
\\

\section{Magnetic moments in the Anti Ferromagnetic case}
\begin{figure}[h!]
\centering
\includegraphics[width=1\linewidth]{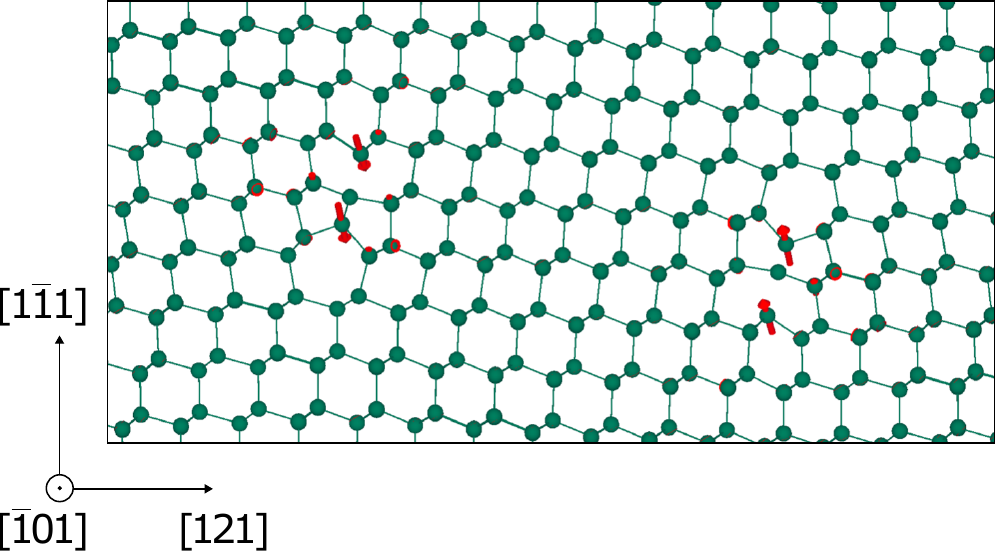}
\caption{Visualization of magnetic moments in the S3-S3 dislocations dipole supercell in the AFM solution}
\label{fig:S3}
\end{figure}

To further support the analysis presented in Fig.\ref{fig:afm}, we explicitly report the real-space distribution of the local magnetic moments within the 60° shuffle S3–S3 dipole supercell, considered in its antiferromagnetic (AFM) configuration. As shown in Fig.\ref{fig:S3}, the magnetic moments are strongly localized at the cores of the two dislocations, clearly identifying the regions where the spin polarization is confined. Importantly, the two dislocation cores exhibit magnetic moments of equal magnitude but opposite sign, thereby providing a direct and unambiguous real-space confirmation of the antiferromagnetic coupling between the defects.

\clearpage
\bibliography{bibl}

@article{RovarisPRM2017,
  title={Three-dimensional SiGe/Si heterostructures: Switching the dislocation sign by substrate under-etching},
  author={Fabrizio Rovaris and Fabio Isa and Riccardo Gatti and Arik Jung and Giovanni Isella and Francesco Montalenti and Hans von K{\"a}nel},
  journal={Physical Review Materials},
  year={2017},
  volume={1},
  pages={073602},
  url={https://api.semanticscholar.org/CorpusID:139657112}
}

@article{MooneyMSER1996,
title = {Strain relaxation and dislocations in SiGe/Si structures},
journal = {Materials Science and Engineering: R: Reports},
volume = {17},
number = {3},
pages = {105-146},
year = {1996},
issn = {0927-796X},
doi = {https://doi.org/10.1016/S0927-796X(96)00192-1},
url = {https://www.sciencedirect.com/science/article/pii/S0927796X96001921},
author = {P.M. Mooney},
}

@article{marzegalliprb2013,
  title={Onset of plastic relaxation in the growth of Ge on Si(001) at low temperatures: Atomic-scale microscopy and dislocation modeling},
  author={Marzegalli, A. and Brunetto, M. and Salvalaglio, M. and Montalenti, F. and Nicotra, G. and Scuderi, M. and Spinella, C. and De Seta, M. and Capellini, G.},
  journal={Physical Review B},
  volume={88},
  number={16},
  pages={165418},
  year={2013},
  doi={10.1103/PhysRevB.88.165418}
}

@article{zhang2013intrinsic,
  title={Intrinsic magnetism of grain boundaries in two-dimensional metal dichalcogenides},
  author={Zhang, Zhuhua and Zou, Xiaolong and Crespi, Vincent H. and Yakobson, Boris I.},
  journal={ACS Nano},
  volume={7},
  number={12},
  pages={10475--10481},
  year={2013},
  doi={10.1021/nn4052887}
}

@book{kittel2005introduction,
  title={Introduction to Solid State Physics},
  author={Kittel, Charles},
  edition={8th},
  year={2005},
  publisher={John Wiley \& Sons},
  address={Hoboken, NJ}
}

@misc{ioffe_ge_database,
  author = {{Ioffe Institute}},
  title = {Band structure and carrier concentration of {Germanium} ({Ge})},
  howpublished = {{NSM} Archive - Physical Properties of Semiconductors},
  url = {https://www.ioffe.ru/SVA/NSM/Semicond/Ge/bandstr.html},
  note = {Accessed: 2026}
}

@article{ArroyoSciptaMat2019,
title = {Effect of thermal annealing on the interface quality of Ge/Si heterostructures},
journal = {Scripta Materialia},
volume = {170},
pages = {52-56},
year = {2019},
issn = {1359-6462},
doi = {https://doi.org/10.1016/j.scriptamat.2019.05.025},
url = {https://www.sciencedirect.com/science/article/pii/S1359646219302933},
author = {Rojas Dasilva Arroyo and Fabio Isa and Giovanni Isella and Rolf Erni and Hans {von Känel} and Pierangelo Gröning and Marta D. Rossell},
}

@article{FitzgerladPssa1999,
author = {Fitzgerald, E. A. and Currie, M. T. and Samavedam, S. B. and Langdo, T. A. and Taraschi, G. and Yang, V. and Leitz, C. W. and Bulsara, M. T.},
title = {Dislocations in Relaxed SiGe/Si Heterostructures},
journal = {physica status solidi (a)},
volume = {171},
number = {1},
pages = {227-238},
doi = {https://doi.org/10.1002/(SICI)1521-396X(199901)171:1<227::AID-PSSA227>3.0.CO;2-Y},
url = {https://onlinelibrary.wiley.com/doi/abs/10.1002/%28SICI%291521-396X%28199901%29171%3A1%3C227%3A%3AAID-PSSA227%3E3.0.CO%3B2-Y},
year = {1999}
}

@article{KochNPJQuantum2025,
	title = {Industrial 300 mm wafer processed spin qubits in natural silicon/silicon-germanium},
	volume = {11},
	copyright = {2025 The Author(s)},
	issn = {2056-6387},
	url = {https://www.nature.com/articles/s41534-025-01016-x},
	doi = {10.1038/s41534-025-01016-x},
	number = {1},
	urldate = {2026-02-02},
	journal = {npj Quantum Information},
	publisher = {Nature Publishing Group},
	author = {Koch, Thomas and Godfrin, Clement and Adam, Viktor and Ferrero, Julian and Schroller, Daniel and Glaeser, Noah and Kubicek, Stefan and Li, Ruoyu and Loo, Roger and Massar, Shana and Simion, George and Wan, Danny and De Greve, Kristiaan and Wernsdorfer, Wolfgang},
	year = {2025},
	pages = {59},
}

@Article{Scalise2024,
author ="Marzegalli, Anna and Montalenti, Francesco and Scalise, Emilio",
title  ="Polytypic quantum wells in Si and Ge: impact of 2D hexagonal inclusions on electronic band structure",
journal  ="Nanoscale Horiz.",
year  ="2024",
volume  ="9",
issue  ="12",
pages  ="2320-2325",
publisher  ="The Royal Society of Chemistry",
doi  ="10.1039/D4NH00355A",
url  ="http://dx.doi.org/10.1039/D4NH00355A"}

@article{rodl2019accurate,
archivePrefix = {arXiv},
arxivId = {1812.01865},
author = {R{\"{o}}dl, Claudia and Furthm{\"{u}}ller, J{\"{u}}rgen and Suckert, Jens Ren{\`{e}} and Armuzza, Valerio and Bechstedt, Friedhelm and Botti, Silvana},
doi = {10.1103/PhysRevMaterials.3.034602},
eprint = {1812.01865},
issn = {24759953},
journal = {Physical Review Materials},
month = {mar},
number = {3},
pages = {034602},
title = {{Accurate electronic and optical properties of hexagonal germanium for optoelectronic applications}},
url = {https://journals.aps.org/prmaterials/pdf/10.1103/PhysRevMaterials.3.034602 https://link.aps.org/doi/10.1103/PhysRevMaterials.3.034602},
volume = {3},
year = {2019}
}

@article{matthews1974defects,
  title={Defects in epitaxial multilayers: {I}. {M}isfit dislocations},
  author={Matthews, J. W. and Blakeslee, A. E.},
  journal={Journal of Crystal Growth},
  volume={27},
  pages={118--125},
  year={1974},
  publisher={Elsevier},
  doi={10.1016/S0022-0248(74)80055-2}
}

@article{people1985calculation,
  title={Calculation of critical layer thickness versus lattice mismatch for {Ge$_x$Si$_{1-x}$}/{Si} strained-layer heterostructures},
  author={People, R. and Bean, J. C.},
  journal={Applied Physics Letters},
  volume={47},
  number={3},
  pages={322--324},
  year={1985},
  publisher={AIP Publishing},
  doi={10.1063/1.96206}
}

@article{Fadaly2021,
author = {Fadaly, Elham M.T. and Marzegalli, Anna and Ren, Yizhen and Sun, Lin and Dijkstra, Alain and {De Matteis}, Diego and Scalise, Emilio and Sarikov, Andrey and {De Luca}, Marta and Rurali, Riccardo and Zardo, Ilaria and Haverkort, Jos E.M. and Botti, Silvana and Miglio, Leo and Bakkers, Erik P.A.M. and Verheijen, Marcel A.},
doi = {10.1021/acs.nanolett.1c00683},
issn = {15306992},
journal = {Nano Letters},
number = {8},
pages = {3619--3625},
pmid = {33843244},
title = {{Unveiling Planar Defects in Hexagonal Group IV Materials}},
volume = {21},
year = {2021}
}

@article{Scalise2020,
archivePrefix = {arXiv},
arxivId = {2001.11826},
author = {Scalise, Emilio and Barbisan, Luca and Sarikov, Andrey and Montalenti, Francesco and Miglio, Leo and Marzegalli, Anna},
doi = {10.1039/D0TC00909A},
eprint = {2001.11826},
issn = {2050-7526},
journal = {Journal of Materials Chemistry C},
month = {jan},
number = {25},
pages = {8380--8392},
title = {{The origin and nature of killer defects in 3C-SiC for power electronic applications by a multiscale atomistic approach}},
url = {http://xlink.rsc.org/?DOI=D0TC00909A http://arxiv.org/abs/2001.11826},
volume = {8},
year = {2020}
}

@article{Zhang2026,
author = {Zhang, Cunzhi and Yu, Victor Wen-zhe and Jin, Yu and Nagura, Jonah and Genlik, Sevim Polat and Ghazisaeidi, Maryam and Galli, Giulia},
doi = {10.1038/s41524-025-01945-3},
issn = {2057-3960},
journal = {npj Computational Materials},
title = {{Towards dislocation-driven quantum interconnects}},
url = {https://doi.org/10.1038/s41524-025-01945-3},
year = {2026}
}

@inproceedings{pizzagalli,
  title={Dislocation cores in silicon: new aspects from numerical simulations},
  author={Pizzagalli, Laurent and Godet, Julien and Gu{\'e}nol{\'e}, Julien and Brochard, Sandrine},
  booktitle={Journal of Physics: Conference Series},
  volume={281},
  number={1},
  pages={012002},
  year={2011},
  organization={IOP Publishing}
}

@article{ivady2019stabilization,
  title={Stabilization of point-defect spin qubits by quantum wells},
  author={Iv{\'a}dy, Viktor and Davidsson, Joel and Delegan, Nazar and Falk, Abram L and Klimov, Paul V and Whiteley, Samuel J and Hruszkewycz, Stephan O and Holt, Martin V and Heremans, F Joseph and Son, Nguyen Tien and others},
  journal={Nature communications},
  volume={10},
  number={1},
  pages={5607},
  year={2019},
  publisher={Nature Publishing Group UK London}
}

@article{barragan2024assessing,
  title={Assessing the potential of perfect screw dislocations in \text{SiC} for solid-state quantum technologies},
  author={Barragan-Yani, Daniel and Wirtz, Ludger},
  journal={Physical Review Research},
  volume={6},
  number={2},
  pages={L022055},
  year={2024},
  publisher={APS}
}

@book{hull2011introduction,
  title={Introduction to dislocations},
  author={Hull, Derek and Bacon, David J},
  volume={37},
  year={2011},
  publisher={Elsevier}
}

@article{kresse1996efficient,
  title={Efficient iterative schemes for ab initio total-energy calculations using a plane-wave basis set},
  author={Kresse, Georg and Furthm{\"u}ller, J{\"u}rgen},
  journal={Physical review B},
  volume={54},
  number={16},
  pages={11169},
  year={1996},
  publisher={APS}
}

@article{kresse1996efficiency,
  title={Efficiency of ab-initio total energy calculations for metals and semiconductors using a plane-wave basis set},
  author={Kresse, Georg and Furthm{\"u}ller, J{\"u}rgen},
  journal={Computational materials science},
  volume={6},
  number={1},
  pages={15--50},
  year={1996},
  publisher={Elsevier}
}

@article{perdew2008restoring,
  title={Restoring the density-gradient expansion for exchange in solids and surfaces},
  author={Perdew, John P and Ruzsinszky, Adrienn and Csonka, G{\'a}bor I and Vydrov, Oleg A and Scuseria, Gustavo E and Constantin, Lucian A and Zhou, Xiaolan and Burke, Kieron},
  journal={Physical review letters},
  volume={100},
  number={13},
  pages={136406},
  year={2008},
  publisher={APS}
}

@article{monkhorst1976special,
  title={Special points for Brillouin-zone integrations},
  author={Monkhorst, Hendrik J and Pack, James D},
  journal={Physical review B},
  volume={13},
  number={12},
  pages={5188},
  year={1976},
  publisher={APS}
}

@article{heyd2003hybrid,
  title={Hybrid functionals based on a screened Coulomb potential},
  author={Heyd, Jochen and Scuseria, Gustavo E and Ernzerhof, Matthias},
  journal={The Journal of chemical physics},
  volume={118},
  number={18},
  pages={8207--8215},
  year={2003},
  publisher={American Institute of Physics}
}

@article{perdew2005prescription,
  title={Prescription for the design and selection of density functional approximations: More constraint satisfaction with fewer fits},
  author={Perdew, John P and Ruzsinszky, Adrienn and Tao, Jianmin and Staroverov, Viktor N and Scuseria, Gustavo E and Csonka, G{\'a}bor I},
  journal={The Journal of chemical physics},
  volume={123},
  number={6},
  year={2005},
  publisher={AIP Publishing}
}

@article{tran2007band,
  title={Band gap calculations with Becke--Johnson exchange potential},
  author={Tran, Fabien and Blaha, Peter and Schwarz, Karlheinz},
  journal={Journal of Physics: Condensed Matter},
  volume={19},
  number={19},
  pages={196208},
  year={2007},
  publisher={IOP Publishing}
}

@article{tran2009accurate,
  title={Accurate Band Gaps of Semiconductors and Insulators with a Semilocal Exchange-Correlation Potential},
  author={Tran, Fabien and Blaha, Peter},
  journal={Physical review letters},
  volume={102},
  number={22},
  pages={226401},
  year={2009},
  publisher={APS}
}

@article{becke2006simple,
  title={A simple effective potential for exchange},
  author={Becke, Axel D and Johnson, Erin R},
  journal={The Journal of chemical physics},
  volume={124},
  number={22},
  year={2006},
  publisher={AIP Publishing}
}

@article{csonka2009assessing,
  title={Assessing the performance of recent density functionals for bulk solids},
  author={Csonka, G{\'a}bor I and Perdew, John P and Ruzsinszky, Adrienn and Philipsen, Pier HT and Leb{\`e}gue, S{\'e}bastien and Paier, Joachim and Vydrov, Oleg A and {\'A}ngy{\'a}n, J{\'a}nos G},
  journal={Physical Review B—Condensed Matter and Materials Physics},
  volume={79},
  number={15},
  pages={155107},
  year={2009},
  publisher={APS}
}

@article{zhang2018performance,
  title={Performance of various density-functional approximations for cohesive properties of 64 bulk solids},
  author={Zhang, Guo-Xu and Reilly, Anthony M and Tkatchenko, Alexandre and Scheffler, Matthias},
  journal={New Journal of Physics},
  volume={20},
  number={6},
  pages={063020},
  year={2018},
  publisher={IOP Publishing}
}

@article{LeGouesPRL1991,
  title = {Anomalous strain relaxation in SiGe thin films and superlattices},
  author = {LeGoues, F. K. and Meyerson, B. S. and Morar, J. F.},
  journal = {Phys. Rev. Lett.},
  volume = {66},
  issue = {22},
  pages = {2903--2906},
  numpages = {0},
  year = {1991},
  month = {Jun},
  publisher = {American Physical Society},
  doi = {10.1103/PhysRevLett.66.2903},
  url = {https://link.aps.org/doi/10.1103/PhysRevLett.66.2903}
}

@article{SkibitzkiPRM2020,
  title = {Reduction of threading dislocation density beyond the saturation limit by optimized reverse grading},
  author = {Skibitzki, Oliver and Zoellner, Marvin H. and Rovaris, Fabrizio and Schubert, Markus Andreas and Yamamoto, Yuji and Persichetti, Luca and Di Gaspare, Luciana and De Seta, Monica and Gatti, Riccardo and Montalenti, Francesco and Capellini, Giovanni},
  journal = {Phys. Rev. Mater.},
  volume = {4},
  issue = {10},
  pages = {103403},
  numpages = {7},
  year = {2020},
  month = {Oct},
  publisher = {American Physical Society},
  doi = {10.1103/PhysRevMaterials.4.103403},
  url = {https://link.aps.org/doi/10.1103/PhysRevMaterials.4.103403}
}

@article{polat2023dislocations,
  title={Dislocations as natural quantum wires in diamond},
  author={Polat Genlik, Sevim and Myers, Roberto C and Ghazisaeidi, Maryam},
  journal={Physical Review Materials},
  volume={7},
  number={2},
  pages={024601},
  year={2023},
  publisher={APS}
}

@book{claeys2009extended,
  title={Extended defects in germanium: Fundamental and technological aspects},
  author={Claeys, Cor and Simoen, Eddy},
  year={2009},
  publisher={Springer}
}

@article{cai2001anisotropic,
  title={Anisotropic elastic interactions of a periodic dislocation array},
  author={Cai, Wei and Bulatov, Vasily V and Chang, Jinpeng and Li, Ju and Yip, Sidney},
  journal={Physical Review Letters},
  volume={86},
  number={25},
  pages={5727},
  year={2001},
  publisher={APS}
}

@article{scappucci2021germanium,
  title={The germanium quantum information route},
  author={Scappucci, Giordano and Kloeffel, Christoph and Zwanenburg, Floris A and Loss, Daniel and Myronov, Maksym and Zhang, Jian-Jun and De Franceschi, Silvano and Katsaros, Georgios and Veldhorst, Menno},
  journal={Nature Reviews Materials},
  volume={6},
  number={10},
  pages={926--943},
  year={2021},
  publisher={Nature Publishing Group UK London}
}

@article{barbisan2022atomic,
  title={Atomic-scale insights on the formation of ordered arrays of edge dislocations in Ge/Si (001) films via molecular dynamics simulations},
  author={Barbisan, Luca and Marzegalli, Anna and Montalenti, Francesco},
  journal={Scientific Reports},
  volume={12},
  number={1},
  pages={3235},
  year={2022},
  publisher={Nature Publishing Group UK London}
}

@article{steele2025superconductivity,
  title={Superconductivity in substitutional Ga-hyperdoped Ge epitaxial thin films},
  author={Steele, Julian A and Strohbeen, Patrick J and Verdi, Carla and Baktash, Ardeshir and Danilenko, Alisa and Chen, Yi-Hsun and van Dijk, Jechiel and Knudsen, Frederik H and Leblanc, Axel and Perconte, David and others},
  journal={Nature Nanotechnology},
  pages={1--7},
  year={2025},
  publisher={Nature Publishing Group UK London}
}

@article{peralta2006spin,
  title={Spin-orbit splittings and energy band gaps calculated with the Heyd-Scuseria-Ernzerhof screened hybrid functional},
  author={Peralta, Juan E and Heyd, Jochen and Scuseria, Gustavo E and Martin, Richard L},
  journal={Physical Review B—Condensed Matter and Materials Physics},
  volume={74},
  number={7},
  pages={073101},
  year={2006},
  publisher={APS}
}

@article{hirth1983theory,
  title={Theory of dislocations},
  author={Hirth, John Price and Lothe, Jens and Mura, Toshio},
  journal={Journal of Applied Mechanics},
  volume={50},
  number={2},
  pages={476--477},
  year={1983},
  publisher={ASME International}
}

@article{hornstra1958dislocations,
  title={Dislocations in the diamond lattice},
  author={Hornstra, J},
  journal={Journal of Physics and Chemistry of Solids},
  volume={5},
  number={1-2},
  pages={129--141},
  year={1958},
  publisher={Elsevier}
}

@article{rubel2014unfolding,
  title={Unfolding the band structure of disordered solids: From bound states to high-mobility Kane fermions},
  author={Rubel, O and Bokhanchuk, A and Ahmed, SJ and Assmann, E},
  journal={Physical Review B},
  volume={90},
  number={11},
  pages={115202},
  year={2014},
  publisher={APS}
}

@article{wang1998majority,
  title={“Majority representation” of alloy electronic states},
  author={Wang, L-W and Bellaiche, L and Wei, S-H and Zunger, A},
  journal={Physical review letters},
  volume={80},
  number={21},
  pages={4725},
  year={1998},
  publisher={APS}
}

@article{krivobok2018two,
  title={Two types of isolated (quantum) emitters related to dislocations in crystalline CdZnTe},
  author={Krivobok, VS and Nikolaev, SN and Chentsov, SI and Onishchenko, EE and Pruchkina, AA and Bagaev, VS and Silina, AA and Smirnova, NA},
  journal={Journal of Luminescence},
  volume={200},
  pages={240--247},
  year={2018},
  publisher={Elsevier}
}

@article{hu2018ubiquitous,
  title={Ubiquitous spin-orbit coupling in a screw dislocation with high spin coherency},
  author={Hu, Lin and Huang, Huaqing and Wang, Zhengfei and Jiang, Wei and Ni, Xiaojuan and Zhou, Yinong and Zielasek, V and Lagally, MG and Huang, Bing and Liu, Feng},
  journal={Physical review letters},
  volume={121},
  number={6},
  pages={066401},
  year={2018},
  publisher={APS}
}

@article{naumov2023one,
  title={One-dimensional magnetism and Rashba-like effects in zigzag bismuth nanoribbons},
  author={Naumov, Ivan I and Dev, Pratibha},
  journal={Physical Review Materials},
  volume={7},
  number={2},
  pages={026204},
  year={2023},
  publisher={APS}
}

@article{li2023dislocation,
  title={Dislocation-tuned ferroelectricity and ferromagnetism of the BiFeO3/SrRuO3 interface},
  author={Li, Xiaomei and Han, Bo and Zhu, Ruixue and Shi, Ruochen and Wu, Mei and Sun, Yuanwei and Li, Yuehui and Liu, Bingyao and Wang, Lifen and Zhang, Jingmin and others},
  journal={Proceedings of the National Academy of Sciences},
  volume={120},
  number={13},
  pages={e2213650120},
  year={2023},
  publisher={National Academy of Sciences}
}

@article{liang2023field,
  title={Field-free spin-orbit switching of perpendicular magnetization enabled by dislocation-induced in-plane symmetry breaking},
  author={Liang, Yuhan and Yi, Di and Nan, Tianxiang and Liu, Shengsheng and Zhao, Le and Zhang, Yujun and Chen, Hetian and Xu, Teng and Dai, Minyi and Hu, Jia-Mian and others},
  journal={Nature Communications},
  volume={14},
  number={1},
  pages={5458},
  year={2023},
  publisher={Nature Publishing Group UK London}
}

@article{humble1978plastic,
  title={Plastic deformation of diamond at room temperature},
  author={Humble, P and Hannink, RHJ},
  journal={Nature},
  volume={273},
  number={5657},
  pages={37--39},
  year={1978},
  publisher={Nature Publishing Group UK London}
}

@book{fitzgerald1989properties,
  title={The properties, control and elimination of misfit dislocations in semiconductor heterostructures},
  author={Fitzgerald Jr, Eugene Arthur},
  year={1989},
  publisher={Cornell University}
}

@article{arguirov2014luminescence,
  title={Luminescence from germanium and germanium on silicon},
  author={Arguirov, Tzanimir and Kittler, Martin and Oehme, Michael and Abrosimov, Nikolay V and Vyvenko, Oleg F and Kasper, Erich and Schulze, J{\"o}rg},
  journal={Solid State Phenomena},
  volume={205},
  pages={383--393},
  year={2014},
  publisher={Trans Tech Publ}
}

@article{kittler2011photoluminescence,
  title={Photoluminescence study of Ge containing crystal defects},
  author={Kittler, Martin and Arguirov, Tzanimir and Oehme, Michael and Yamamoto, Yuji and Tillack, Bernd and Abrosimov, Nikolai V},
  journal={physica status solidi (a)},
  volume={208},
  number={4},
  pages={754--759},
  year={2011},
  publisher={Wiley Online Library}
}

@article{pezzoli2014ge,
  title={Ge crystals on Si show their light},
  author={Pezzoli, Fabio and Isa, Fabio and Isella, Giovanni and Falub, Claudiu V and Kreiliger, Thomas and Salvalaglio, M and Bergamaschini, R and Grilli, E and Guzzi, M and von K{\"a}nel, Hans and others},
  journal={Physical Review Applied},
  volume={1},
  number={4},
  pages={044005},
  year={2014},
  publisher={APS}
}

@article{tanaka1996photoluminescence,
  title={Photoluminescence spectra of deformed Si-Ge alloy},
  author={Tanaka, K and Suezawa, M and Yonenaga, I},
  journal={Journal of applied physics},
  volume={80},
  number={12},
  pages={6991--6996},
  year={1996},
  publisher={American Institute of Physics}
}

@article{SHIMURA2024108231,
title = {Compressively strained epitaxial Ge layers for quantum computing applications},
author = {Yosuke Shimura and Clement Godfrin and Andriy Hikavyy and Roy Li and Juan Aguilera and Georgios Katsaros and Paola Favia and Han Han and Danny Wan and Kristiaan {De Greve} and Roger Loo},
journal = {Materials Science in Semiconductor Processing},
volume = {174},
pages = {108231},
year = {2024},
publisher={Elsevier}
}

@article{WANG20011599,
title = {Dislocation evolution in epitaxial multilayers and graded composition buffers},
author = {T.C Wang and Y.W Zhang and S.J Chua},
journal = {Acta Materialia},
volume = {49},
number = {9},
pages = {1599-1605},
year = {2001},
publisher={Elsevier}
}

@article{Bolkhovityanov_2012,
title = {Ge-on-Si films obtained by epitaxial growing: edge dislocations and their participation in plastic relaxation},
author = {Bolkhovityanov, Yu B and Sokolov, L V},
journal = {Semiconductor Science and Technology},
volume = {27},
number = {4},
pages = {043001},
year = {2012},
publisher = {IOP Publishing},
}

@article{PhysRevApplied.10.054067,
  title = {Misfit-Dislocation Distributions in Heteroepitaxy: From Mesoscale Measurements to Individual Defects and Back},
  author = {Rovaris, Fabrizio and Zoellner, Marvin H. and Zaumseil, Peter and Schubert, Markus A. and Marzegalli, Anna and Di Gaspare, Luciana and De Seta, Monica and Schroeder, Thomas and Storck, Peter and Schwalb, Georg and Richter, Carsten and Sch\"ulli, Tobias U. and Capellini, Giovanni and Montalenti, Francesco},
  journal = {Phys. Rev. Appl.},
  volume = {10},
  issue = {5},
  pages = {054067},
  numpages = {7},
  year = {2018},
  month = {Nov},
  publisher = {American Physical Society},
  doi = {10.1103/PhysRevApplied.10.054067},
  url = {https://link.aps.org/doi/10.1103/PhysRevApplied.10.054067}
}

@article{ogawa2011ab,
  title={An ab initio analysis of electronic states associated with a silicon vacancy in cubic symmetry},
  author={Ogawa, T and Tsuruta, K and Iyetomi, H},
  journal={Solid state communications},
  volume={151},
  number={21},
  pages={1605--1608},
  year={2011},
  publisher={Elsevier}
}

@article{atta2004systematic,
  title={Systematic study of electron localization in an amorphous semiconductor},
  author={Atta-Fynn, Raymond and Biswas, Parthapratim and Ordej{\'o}n, Pablo and Drabold, DA},
  journal={Physical Review B},
  volume={69},
  number={8},
  pages={085207},
  year={2004},
  publisher={APS}
}

@article{wilhelmer2023over,
  title={Over-and undercoordinated atoms as a source of electron and hole traps in amorphous silicon nitride (a-Si3N4)},
  author={Wilhelmer, Christoph and Waldhoer, Dominic and Cvitkovich, Lukas and Milardovich, Diego and Waltl, Michael and Grasser, Tibor},
  journal={Nanomaterials},
  volume={13},
  number={16},
  pages={2286},
  year={2023},
  publisher={MDPI}
}

@article{grillot1996effect,
  title={Effect of composition on deep levels in heteroepitaxial GexSi1- x layers and evidence for dominant intrinsic recombination-generation in relaxed Ge layers on Si},
  author={Grillot, PN and Ringel, SA and Fitzgerald, EA},
  journal={Journal of electronic materials},
  volume={25},
  number={7},
  pages={1028--1036},
  year={1996},
  publisher={Springer}
}

@article{hubbard1980electrical,
  title={Electrical properties of dislocations in ultra-pure germanium},
  author={Hubbard, G Scott and Haller, Eugene E},
  journal={Journal of Electronic Materials},
  volume={9},
  number={1},
  pages={51--66},
  year={1980},
  publisher={Springer}
}

@dataset{regazzoni_2026_22796437,
  author={Regazzoni, Veronica and Rovaris, Fabrizio and Marzegalli, Anna and Scalise, Emilio},
  title={Atomistic structures of the 60° dislocations in 3C Germanium},
  month=sep,
  year=2026,
  publisher={Zenodo},
  doi={10.5281/zenodo.22796437},
  url={https://doi.org/10.5281/zenodo.22796437},
}

\end{document}